\documentclass[a4paper,12pt]{article}
\usepackage{graphicx}

\begin{document}


\begin{titlepage}

\begin{center}
\Large{\bf Co-expression of statistically over-represented
peptides in proteomes: a key to phylogeny ? }
\end{center}
\vspace{2cm}

\begin{center}
\large{ Luca Ferraro$^{1,\ddag}$, Andrea Giansanti$^{2,3\S}$, 
Giovanni Giuliano$^{4}$ and Vittorio Rosato.$^{1,3}$}
\end{center}
\vspace{1cm}
\begin{center}

\normalsize{ ${^1}$ {Computing and Modelling Unit (CAMO)\\ ENEA
Ente per le Nuove Tecnologie, L' Energia e l' Ambiente\\Casaccia
Research Center, P. O. Box 2400, 00100 Roma, Italy.}

\vspace{.5cm} $^{2}$ {Dipartimento di Fisica, Universit\`a di Roma
``La Sapienza''\\ Piazzale Aldo Moro 2, 00185 Roma, Italy.}

\vspace{.5cm}
${^3}$ {INFM, Unit\`a di Ricerca di Roma1\\ Piazzale Aldo Moro 2,
00185 Roma, Italy.}
\vspace{.5cm}

${^4}$ {Biotechnology Unit (BIOTEC)\\ ENEA Ente per le Nuove Tecnologie,
L' Energia e l' Ambiente\\Casaccia Research Center, P. O. Box 2400, 00100 Roma,
Italy.}
\vspace{.5cm}

}
\end{center}
\vspace{.5cm}

$^{\S}$Corresponding author\\

$^\ddag$ present address:
\noindent
CASPUR, Consorzio Interuniversitario per le Applicazioni di
Supercalcolo per Universit\`a e Ricerca, via dei Tizii 6/b, 00185 Roma,
Italy.

\end{titlepage}

\pagestyle{empty}
\begin{center}
{\bf Abstract}
\end{center}
\vspace{0.5cm}

It is proposed that the co-expression of statistically significant
motifs among the sequences of a proteome is a phylogenetic trait.
From the co-expression matrix of such motifs in a group of
prokaryotic proteomes a suitable definition of a phylogenetic
distance is introduced and the corresponding distance matrix
between proteomes is constructed. From the distance matrix a
phylogenetic tree is inferred, following a standard procedure. 
The inferred tree is compared with with a reference tree 
deduced from a distance matrix obtained from the alignment 
of ribosomal RNA sequences. 
Our results are consistent with the hypothesis that biological
evolution manifests itself with a modulation of basic correlations
between shared peptides of short length, present in protein
sequences. Moreover, the simple procedure we propose confirms
that it is possible, sampling entire proteomes, to average the
effects of lateral gene transfer and infer reasonable phylogenies.

\vspace{1cm}
\newpage
\pagestyle{plain}
\addtocounter{page}{-1}


Each living species is the result of its evolution; this
historical assumption is the basic tenet of modern evolutionary
biology. Molecular systematics \cite {Molsis96} aims at
classifying living species by measuring differences in their
inherited molecular constituents, not in their phenotypic,
macroscopic appearance. Since the classic paper by Zuckerkandl and
Pauling \cite{ZuPa65}, rational molecular systematics rests on the
analysis of actual sequences, which are, though in an indirect
way, the archive of the evolutionary information about biological
species (taxa). Differences in nucleotide or amino acid sequences
are the objective material elements to start from; that is
particularly true in the classification of microscopic,
unicellular organisms, where more macroscopic methods, based on
ecological or pathological properties of these species, seem to be
less fundamental \cite{OlWoeOv94}. Carl Woese, more than twenty
years ago, has founded the universal molecular classification of
living organisms based on ubiquitary co-evolved sequences, like
e.g. the RNA of the small ribosomal subunit (SSU
rRNA) \cite{WoeFox77}. The main achievement was the discovery of
the fundamental tripartition of the tree-of-life into the branches
of Bacteria, Archea and Eukarya \cite{doolittle}. Nowadays
molecular phylogeny is a well established discipline, based on
probabilistic methods \cite{WheLiGo01}; nevertheless, the
existence of lateral gene transfer \cite{Doolittle2}\cite{brown0} (i. e. mixing
up of genes between species, particularly practiced among

prokaryotes) has created some problems in the field and led some radical
phylogenists to argue against the reliability of single-gene
phylogenies \cite{Phil98}. To cope with this problem there has
been, since more than ten years ago, a growth of ideas and methods
to infer molecular phylogenies not from the analysis of groups of
single genes, as classically done, but rather from the analysis of
whole genomes and proteomes \cite{moltagente}. This is the subject
also of the present work.

Proteomes are far from being a random assembly of peptides.
Clustering of aminoacids \cite{Rosato}, and strong correlations
among  genomic \cite{Siggia} and proteomic \cite{RigFlo} 
segments have been clearly demonstrated. These results give
meaning to the metaphor of protein sequences viewed as texts
written in a still unknown language \cite{Searls}.

Following this view we assume that biological evolution could
manifest itself, at the molecular level, through the modulation of
{\it significant} sequence elements that can be variously combined
in the evolutionary declination of the language embodied by the
proteins of living organisms. It is reasonable
 to consider as {\it significant} those tracts of a protein sequence which
exhibit a pronounced deviation from a random assembly of
aminoacids. We have then looked for putatively significant
elements as short peptide sequences of lenght $k$ which occur, in
a proteome, a number of times larger than expected in a random
proteome. In this work we show that these peptides are acted
upon by natural selection and display, in different proteomes,
statistical correlations able to express evolutionary distances.

A proteome $P$ is a collection of $n_P$ protein sequences, i.e.
strings of various lengths made of symbols from an alphabet $A$ of
20 letters: $A=\{\sigma_{1},\sigma_{2},...,\sigma_{20}\}$; each
$\sigma$ labels one of the different aminoacids a protein is made
of. If the proteome contains $N_P$ aminoacids (i. e. letters),
then one can compute their relative frequencies:
$f(\sigma_{i})=n_{i}/N_{P}$, $n_{i}$ being the number of times the
$i$-th aminoacid occurs in the proteome and $i=1,2,...,20$. We
define as $k$-peptides sequences of $k$ contiguous letters. The
number of all possible $k$-peptides is $20^k$. From the proteome
$P$ we can only select $N_{P}-n_P\cdot(k-1)$ overlapping $k$-peptides; some
of them occur once, others  more than once, doing so in the same or 
in different different proteins.

 Denoting with
$p^{(k)}_{j}=\{\sigma_{1_{j}},\sigma_{2_{j}},...,\sigma_{k_{j}}\}$
the $j$-th $k$-peptide we can count the number $N^{(o)}_{j}$ of
times it occurs in the actual proteome.
We can also estimate the {\it expected} number of occurrences 
$N^{(e)}_{j}$ of the $j$-th $k$-peptide in a random proteome 
(of the same length and with the same number of $k$-peptides),
generated by independent random extractions of letters with 
the constraint of producing, on the average, prescribed relative 
frequencies $f(\sigma_{i})$. That is:

\begin{equation}
N^{(e)}_{j}=[(N_{P}-n_P\cdot(k-1)] \cdot \Pr[p^{(k)}_{j}],
\end{equation}
\noindent
where $\Pr[p^{(k)}_{j}]$, the probability of occurrence of the 
$j$-th $k$-peptide, can be estimated as $f(\sigma_{1_{j}}) 
\cdot f(\sigma_{2_{j}}) \cdot...\cdot f(\sigma_{k_{j}})$, i. e. as the 
product of the relative frequencies of its component letters
in the actual proteome.

For each $k$-peptide of expected occurrence $N^{(e)}$, the
probability that it is observed $N$ times in a random proteome (with the 
same amino acidic composition and sequences of the same lengths
as in the actual proteome) is given by a Poissonian distribution:

\begin{equation}
Pr _{N^{(e)}}[N]=\frac{{[N^{(e)}]}^{N}} {N!}  \cdot
\exp{[-N^{(e)}]}.
\end{equation}

We define as {\it statistically-relevant} the over-expressed
$k$-peptides whose observed number of occurrences $N^{(o)}$ is such that:

\begin{equation}
\int_{0}^{N^{(o)}}  Pr_{N^{(e)}}[N']dN' \geq 0.95
\end{equation}
\noindent
(i.e. the observed occurrence of the $k$-peptide falls in the
upper five percent tail of its Poissonian distribution). Let us
call hereafter {\it $k$-motifs} the over-expressed $k$-peptides,
selected following inequality (3). Analogously, we have defined a
test-set of $k$-peptides which, differently from $k$-motifs, are
expressed as expected (i.e. $N^{(o)}\approx N^{(e)}$). We call these
peptides \emph{expected} $k$-peptides. $k$-motifs and expected
$k$-peptides are differently distributed along protein sequences:
$k$-motifs are seldom alone in a protein and in many cases they
partly overlap forming longer, potentially significant, tracts.
They occur at specific distances one from the other, whereas
expected $k$-peptides are isolated and dispersed, without any
recurrent clustering. As an example, in fig.1 we show the occurrence
of $k$-motifs in an archaeal protein.

The non trivial statistical properties of the $k$-motifs suggest
that, among the $k$-peptides present in a proteome, they could
display patterns of correlated expression useful to derive
phylogenetic distances between taxa.

We have considered eighteen proteomes from the Gene Bank 
\cite{genebank}. Ten from Archaea {\it Aeropyrum pernix,
Archaeoglobus fulgidus, Halobacterium spNRC1, Methanococcus
jannaschii, Methanobacterium thermoautotrophicum, Pyrobaculum
aerophilum, Pyrococcus abyssi, Pyrococcus furiosus, Sulfolobus
solfataricus, Thermoplasma acidophilum} and eight from Bacteria:
{\it Agrobacterium tumefaciens, Bacillus subtilis, \\Chlorobium
tepidum, Deinococcus radiodurans, Escherichia coli K12,
Synechocystis spPCC6803, Thermotoga maritima, Yersinia pestis CO92
}. We have selected $k$-motifs from all these proteomes and
collected them into {\it $k$-dictionaries}. Let $Z_{n}(k)$ be the subset
of the $k$-dictionary composed by those $k$-motifs which are
expressed in, {\it at least}, $n$ different proteomes. $Z_{1}(k)$
is thus the entire set of $k$-motifs, referring to the considered
group of proteomes. $Z_{2}(k)$ is the set of $k$-motifs common to
{\it at least} two proteomes; $Z_{1}(k)-Z_{2}(k)$ is, therefore,
the subset of the $k$-motifs specific to one proteome.
Fig. 2 reports, as a function of $k$, the number of
entries in different $k$-dictionaries, normalized over the total
number of expressed $k$-peptides ($Z_{1}(k)$). It is worth noting that, 
as $k$ increases, the proteome-specific $k$-motifs 
($Z_{1}(k)-Z_{2}(k)$) rapidly overwhelm
the shared $k$-motifs (i.e. $Z_{6}(k)$). 
The $Z_{2}(k)$ dictionary (open circles in fig.2)
has a significant number of entries for low and intermediate $k$
values, as it contains almost $10\%$ of the expressed
peptides, for $k=6$.

We define now the \emph{co-expression} matrix of the set $
Z_{n}(k)$ in a proteome $P$ the matrix $A^{(P)}[Z_{n}(k)]$; its element $ij$
counts the number of times $i$-th and $j$-th $k$-motifs, from $
Z_{n}(k)$, occur together in one of the proteins of $P$. This
matrix resembles the adjacency matrix of the network formed by
linking words when they occur in the same phrase in texts written
in natural languages, as done in recent linguistic studies 
\cite{Sole}. In a subsequent more extended paper we shall present
the statistical properties of the linguistic co-expression networks  
built on sets of $k$-motifs \cite{Rosato1}.

The pattern of co-expression matrices based on $ Z_{n}(k)$, for a
given value of $k$, is far from trivial in all the considered
proteomes, with many groups of $k$-motifs co-expressed in one or
more proteins up to several tens of times. On the other hand we have 
noticed that co-expression matrices
generated by equally-populated sets of expected $k$-peptides are
sparse, with just very few and tiny elements different from zero.

The different co-expression patterns of $k$-motifs in different
proteomes are the basis of the method we propose in this letter.

The observations reported above might be resumed as follows: (1)
there is a consistent set of $k$-motifs which are common among the
considered organisms of a given kingdom; this set might constitute
a sort of basic dictionary collecting robust pieces of
information, stable across the taxa; (2) there is a larger set of
proteome-specific $k$-motifs [$ Z_{1}(k) - Z_{2}(k)$], whose
evolution occurred within a specific taxon and might be considered
as the manifestation of a {\it linguistic} specificity of that
species; (3) there is a consistent set of $k$-motifs, $ Z_{2}(k)$,
containing the common $k$-motifs together with a number of
$k$-motifs which are quite specific but, nevertheless, common to a
few species.

It is reasonable to assume that common and proteome-specific
$k$-motifs somehow interact: the usage of proteome-specific terms
might influence the usage of the common $k$-motifs, in the sense
that the co-expression of the latter might be modulated by usage
of the former, giving origin to a specific co-expression pattern.

Let us propose now a definition of the phylogenetic distance among
proteomes. From $ A^{(P)}[Z_{n}(k)]$, the symmetric co-expression
matrix of a given proteome $P$, we can extract a co-expression
vector $V^{(P)}[Z_{n}(k)]$, whose components are the
$n_{k}(n_{k}+1)/2$ distinct entries of the matrix ($n_{k}$ is the
number of $k$-motifs in $Z_{n}(k)$), ordered in an arbitrary but
fixed way, e. g. by rows:

\begin{equation}
V^{(P)}_{s}[Z_{n}(k)]=  A^{(P)}_{ij}[(Z_{n}(k)]
\end{equation}
\noindent
with $j\geq i$ and $s$ ranging from one to $n_{k}(n_{k}+1)/2$. We
consider the co-expression vector as a {\it linguistic}
fingerprint of a proteome expressing its peculiar use of both common
and proteome-specific motifs. We define a phylogenetic distance
 $d_{P'P''}(k)$ between two proteomes $P'$ and $P''$ through the
scalar product\cite{note} of their co-expression vectors 
based on a $Z_{j}(k)$ dictionary:

\begin{equation}
d_{P'P''}(j,k) = 1 - \sum _{s}\{ V^{(P')}_{s}[Z_{n}(k)]\cdot
V^{(P'')}_{s}[Z_{n}(k)] \} / \{|\bf V^{(P')}| \cdot |\bf
V^{(P'')}|\}
\end{equation}

In this work we have evaluated phylogenetic distances among a set of
 prokaryotic proteomes, using the $Z_{2}(6)$ dictionary.
There are several arguments to motivate this choice for the probe-set
of motifs. The $Z_{2}(6)$ dictionary of the set of prokaryotes we 
are considering has 7712 entries; it contains a balanced mixture of 
common and proteome-specific tracts. The use of a $Z_{j}(6)$ 
dictionary, with $j>2$ would have produced a distance evaluation 
on the only basis of strongly conserved
motifs, i.e. those common to a large number of organisms,
disregarding the modulation effect that they could produce on the
proteome-specific tracts. These dictionaries, moreover, are quite
small (the $Z_{6}(6)$ dictionary, for instance, contains only 55
entries) and their size markedly decreases with the increase of
both $j$ and $k$.
On the other side, $Z_{2}(k)$ dictionaries with $k<6$ are made
by a large number of motifs (e. g. $Z_{2}(5)$ has 161903 entries),
but the potential increase in sensitivity, putting aside practical
considerations due to the treatment of large matrices, would be
spoiled by some volatility of low $k$ motifs. Due to the rigidity
of the statistical criterion (3) (same acceptance threshold for
all the motifs) one $k$-motif which has passed the test and
belongs to the $k$-dictionary could pass the test for the $k+1$
dictionary as part of one of the 40 ($k+1$)-peptides which can be
obtained by adding one letter at its beginning or at its end. We
have observed that this is rarely the case for low $k$. So, if $k$
is too low then many short peptides, which are accepted as statistically
significant and could be the nucleus of biologically relevant
tracts of sequence, are lost and not recognized in the $k+1$ test.
When $k$ is larger than 5, the motifs have been seen to be more
stable, in that they generally appear as part of longer motifs
also in the $k+1$ dictionary. Indeed, some of them are also "lost" 
but this can be the sign of the "end" of the specific tract. The
$Z_{2}(6)$ dictionary is thus a good trade-off between number of
entries and balance of common and proteome-specific tracts.
Moreover, $k=6$ seem to be a peculiar length for peptides:
it has been proven that $6$-peptides allow a unique
reconstruction  of a protein sequence from the collection of its
constituent $k$-peptides \cite{Huo}.

By using the definition of eq.(5) we have evaluated all the
distances between the considered set of proteomes. The resulting
distance matrix has been processed by the neighbor-joining method 
\cite{Saitou_Nei} using the PHYLIP package \cite{phylip}. The
dendrogram we have obtained through the procedure outlined above
is shown in Fig. 3. In Fig. 4 we show the tree obtained, for the
same set of taxa, from the server of the Ribosomal Database
Project \cite{rdp}. This last phylogeny can be assumed as a reference,
 because it is based on the alignment of sequences of RNA
from the Small Ribosomal Subunit. This molecule is ubiquitary and
coevolved to accommodate a well defined set of ribosomal proteins,
hardly subject to lateral gene transfer. The tree from alignment
of the SSUrRNA shows the clear separation of the two kingdoms:
Archaea and Bacteria, this separation appear less clearly resolved
in our tree whose center seems to be an archaeal spot from which emerge
{\it Sulfolobus solfataricus}, a branch of 5 bacteria, a group of Archaea with
the bacterium {\it Deinococcus radiodurans} (D. radi) among them, and a group 
of Archaea with the bacteria {\it Chlorobium tepidum} (C.tepi) 
and {\it Synechocystis} (Synech) 
segregated among them. One could be discouraged by this result 
and think that the method we are 
proposing is unable to resolve the basic tripartition of the 
tree of life and that 
we are mistakenly classifying taxa. 
One could argue, from a different perspective, that
the kind of method we are proposing, based on global statistical properties 
of the proteomes, is able to reveal phylogenetic associations which are at
variance with the fundamental SSUrRNA classification. The stability of
the method and its biological foundations have to be further investigated.
However it is worth noting that, quite surprisingly, the tree we have 
reconstructed
through a ¨ biologically blind¨ criterion  mutuated from 
statistical linguistics
can be reasonably compared with those obtained through refined and 
deep whole-genome analyses \cite{brown}\cite{fitz-gibbon}.
In particular we believe that whole-genome phylogenies of the kind we are
proposing should be confronted with very recent observations suggesting that
eukaryotes could originate from the fusion of pre-existing prokaryotic genomes 
\cite{lake1}. Moreover, the important distinction between {\it operational}
 and {\it informational} genes \cite{lake2} suggests that we are looking at a 
possible different statistics of occurrence of the $k$-motifs, 
which are the probe of our method, over the two kinds of proteins; 
we also believe that blind
approaches based on the statistics of short sequence motifs, 
as the one we present here,
could be less affected by different sources of bias which are 
however present in
statistical phylogenomic studies based on the clustering of 
entire genes \cite{driskell}. 

In the last stage of the preparation of this manuscript we became
aware of an important study which uses an approach very close to ours  
\cite{bailin1} and which has been made available on line\cite{bailin2}. 
In that method a proteome
is also sampled for statistically significant $6$-peptides; the 
background constituted by an uncorrelated random extraction of letters
is subtracted. The fingerprint vector of each proteome has $20^6$ 
components, each one of them expresses the statistical
deviation of the occurrence of each peptide from that expected in
a random proteome. In our approach the fingerprint is represented 
instead by the co-expression vector.
Following the method proposed in \cite{bailin2} we have 
derived a distance matrix of the same 18 species here 
investigated; the phylogenetic tree we have obtained has a more 
resolved dichotomy between 
Archaea and Bacteria, and a topology which, 
though more consistent is still less resolved and definitely 
not coincident with that of a tree obtained from the 
distance matrix based on the alignment 
of the SSUrRNA \cite{rdp}. 
It will be interesting to proceed, in the next future, 
to a careful assessment of the biological information which can be 
derived from the two approaches. 
At present we tend to have the following view: the phylogenetic picture 
based on the tree of life has been put under scrutiny by the
large extent of lateral gene transfer between taxa; that challenged 
phylogenists, using properly selected groups of genes, to reveal 
evolutionary  relations which are not consistent with the universal 
tree of life. Recently there have been claims for the tree of life
to fuse into what has been called the {\it ring of life} \cite{lake1}.

Phylogenies based on whole genomes are coherent with the 
view of the three kingdoms Archaea, Bacteria and Eukarya as 
originating from a world based on gene exchange and fusion of genomes. 
In particular, testing entire proteomes 
against patterns of correlated expression of statistically significant 
sequence motifs seems to be a proper way to cope with the original 
genome fusion regime and with the
{\it mean field} generated by lateral gene transfers, 
gene duplication and lost. The method proposed in \cite{bailin1}, 
samples in a more generic way the evolutionary correlations between $k$-motifs
and seem to force the trees toward the tree of life shape. Our method, based on
patterns of co-expression of $k$-peptides could be more 
in agreement with the view of a fusion-based ring of life. Of course,
quantitative comparison between different methods is now really required, 
we are planning an extensive quantitative investigation of the 
relative merits of different approaches in recontructing 
the philogeny(ies) of a properly selected set of taxa. 
In doing that a clear mathematical setting is of tantamount 
importance \cite{billera}\cite {diaconis}.

The scientific content of phylogenies that are based on 
the statistical sampling of entire proteomes and that avoid
sequence alignment algorithms has still to be validated.  
Nevertheless we believe that they can have a practical 
relevance at least as tools for the rapid molecular classification 
of the ever increasing number of freshly sequenced genomes.

\section*{Acknowledgments}

The authors have benefited of illuminating discussions with M.
Helmer Citterich, A. Via, A. Zanzoni and G. Ausiello at the
University of Roma "Tor Vergata" and with P. Cammarano at the
University of Roma "La Sapienza".

\newpage

\begin{figure}
\begin{center}
 \centerline { }
 \includegraphics[width=6in]{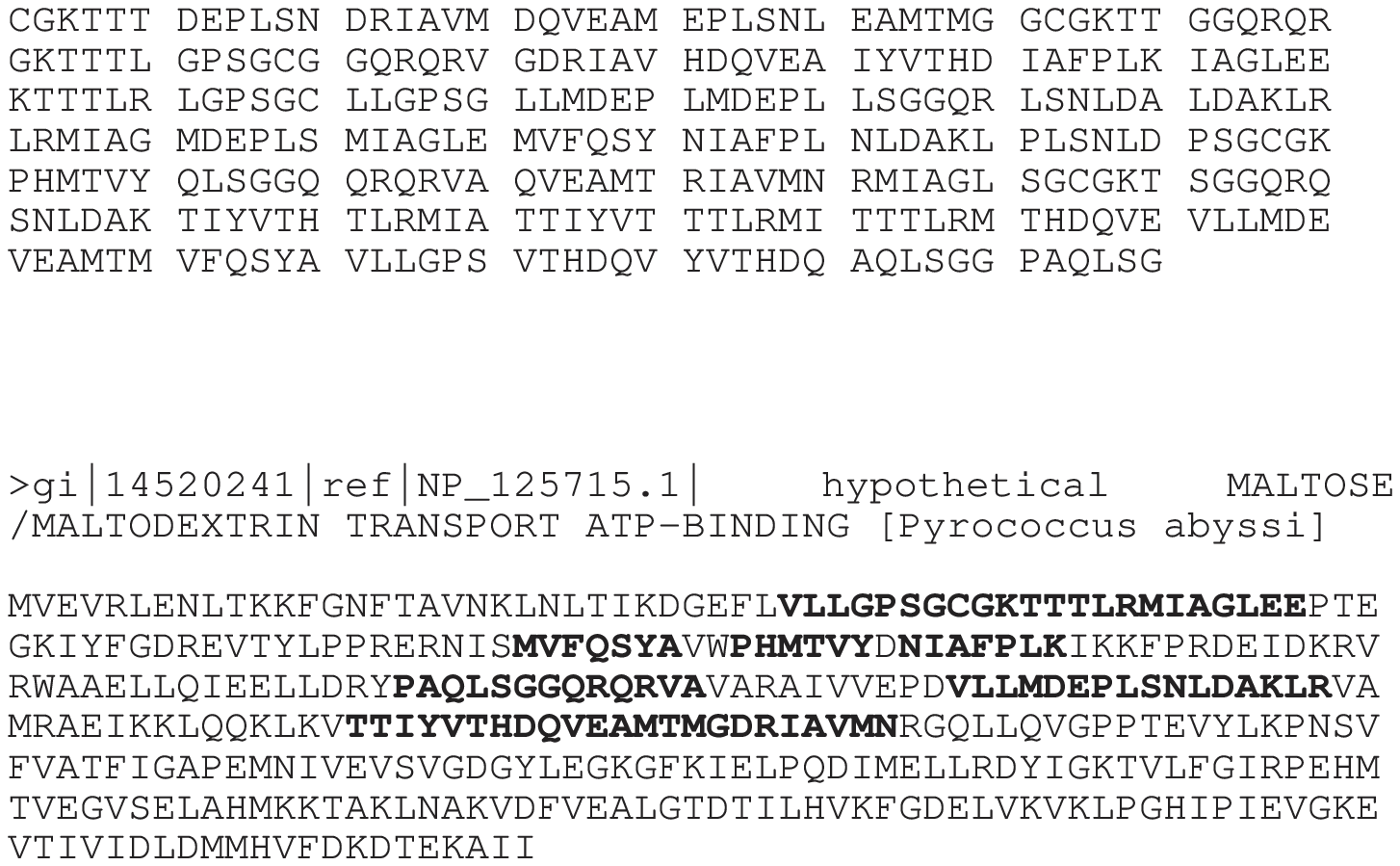}
\end{center}
\caption{A typical dispersion of $6$-motifs (bold) in a
protein of the archaeon {\it P. abyssi}. In the upper part of the figure are
reported the 55 $6$-motifs, belonging to $Z_2(6)$, which are expressed in the
protein. Note the clustering and overlap of the motifs.}
\end{figure}

\begin{figure}
\begin{center}
 \centerline { }
 \includegraphics[width=6in]{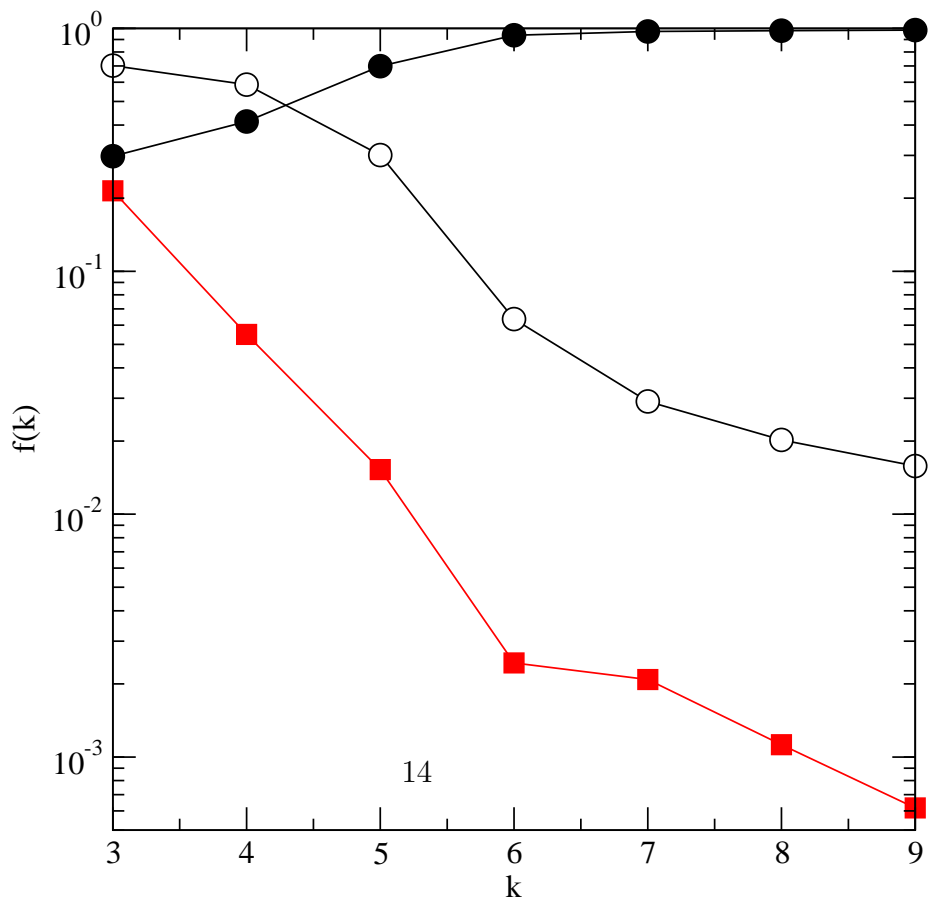}
\end{center}
\caption{Relative fraction of $k$-motifs present in the
different subsets of the k-dictionary: $ Z_{1}(k)-Z_{2}(k)$ (black
circles),$ Z_{2}(k)$ (white circles), $ Z_{6}(k)$ (squares).}
\end{figure}

\begin{figure}
\begin{center}
 \centerline { }
 \includegraphics[width=5in]{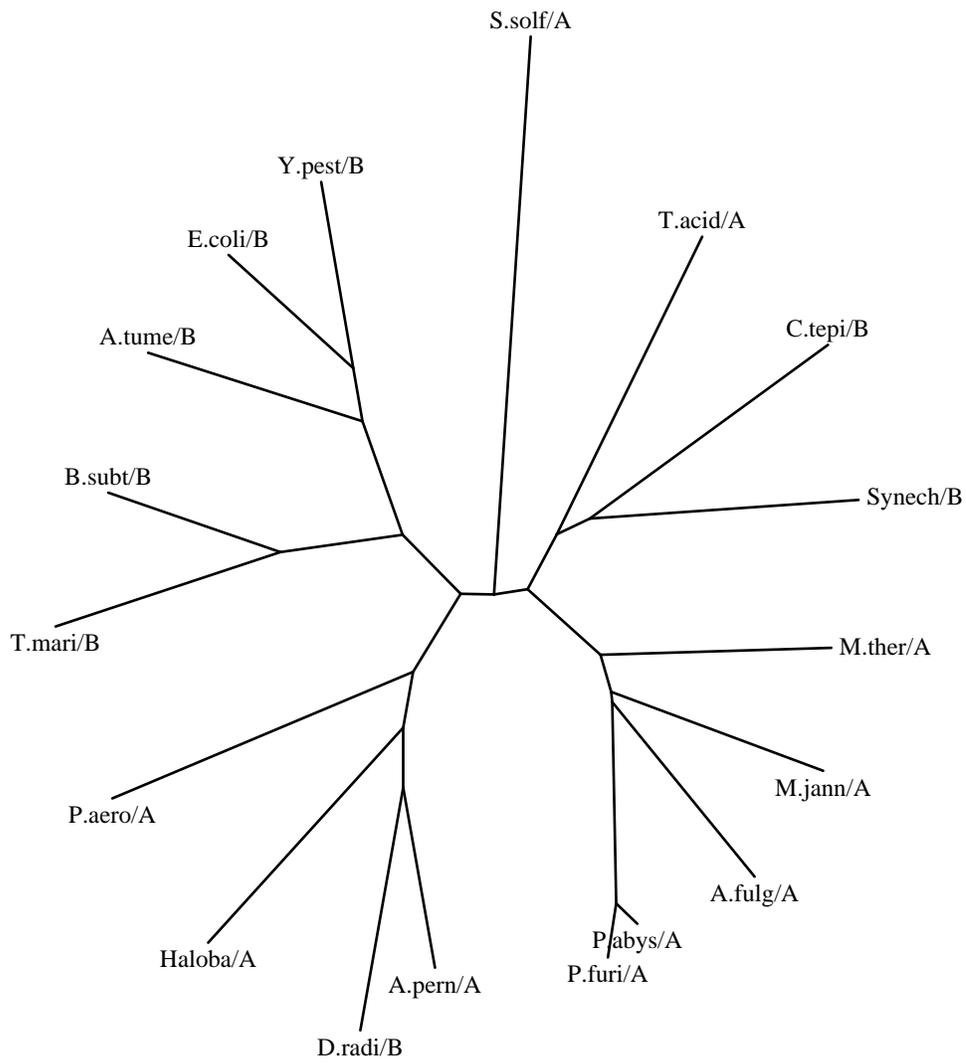}
\end{center}
\caption{Unrooted phylogenetic tree of the considered proteomes: the full name 
of the species can be easily reconstructed from the abbreviations. 
After a / the kingdom is indicated: A stands for Archaea, B for Bacteria.}
\end{figure}

\begin{figure}
\begin{center}
 \centerline { }
 \includegraphics[width=6in]{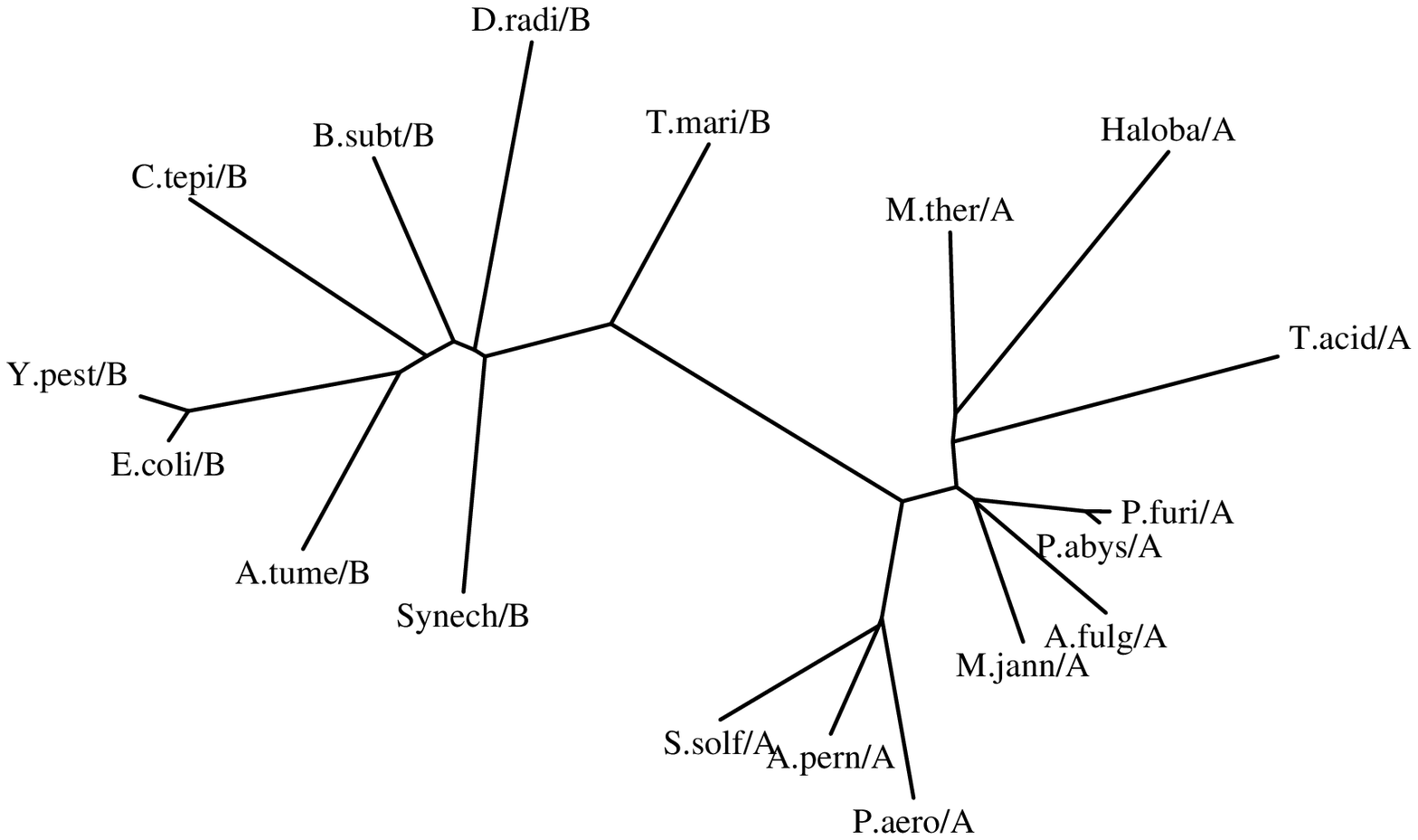}
\end{center}
\caption{SSUrRNA phylogeny of the 18 species here considered; from \cite{rdp}.}
\end{figure}


\begin{thebibliography}{00}


\bibitem{Molsis96} D. M. Hillis, C. Morris and B. K. Mable, {\it Molecular Systematics}, Sinauer Associates, Inc. Sunderland, MA, USA (1996).

\bibitem{ZuPa65} E. Zuckerkandl and L. Pauling, {\it Molecules as documents of evolutionary history}. {Journal of Theoretical
Biology } {\bf 8}:357-366 (1963).

\bibitem{OlWoeOv94} G. J. Olsen, C. R. Woese and R. Overbeek,
{\it The Winds of (Evolutionary) Change: Breathing New Life into

Microbiology }. { Journal of Bacteriology} {\bf 176}: 1-6 (1994).
See also:

\bibitem{WoeFox77} C. R. Woese and G. E. Fox,
{\it Phylogenetic structure of the prokaryotic domain: the primary
kingdoms}. { Proc. Natl. Acad. Sci. USA} {\bf 74}:5088-5090 (1977).

\bibitem{doolittle} W. F. Doolittle, {\it Phylogenetic Classification
and the Universal Tree }. {Science} {\bf 284}:2124-2128 (1999).
And : http://tolweb.org/tree/phylogeny.html.

\bibitem{WheLiGo01} S. Whelan, P. Li\`o and N. Goldman.
 {\it Molecular Phylogenetics:  state-of-the-art methods for
 looking into the past}. {Trends. Genet.}
{\bf 17}: 262-272 (2001).

\bibitem{Doolittle2} W. F. Doolittle, {\it Lateral Genomics}.
{Trends Cell Biol.} {\bf 9}:M5­8 (1999). See also: A. M. Campbell,
{\it Lateral gene
transfer in prokaryotes }. {Theor. Population Biology} {\bf 57}:71-77 (2000).

\bibitem{brown0} J. R. Brown, {\it Ancient horizontal gene transfer}. 
{Nature Reviews Genetics} {\bf 4}: 121-132 (2003).

\bibitem{Phil98} H. Philippe and J. Laurent, 
{\it How good are deep phylogenetic
trees? }.{Curr. Opin. Genet. Devel.} {\bf 8}:616-623 (1998).

\bibitem{moltagente} F. Tekaia, A. Lazcano and B. Dujon, {\it The
Genomic tree as revealed from whole proteome comparisons}.
{Genome Res.}{\bf 9}:550-557 (1999). More recent: G. W. Stuart and
M. W. Berry, {\it A Comprehensive whole genome bacterial phylogeny
using correlated peptide motifs defined in a high dimensional
vector space}.{Journal of Bioinformatics and Computational
Biology} {\bf 1}:475-493 (2003) and references 1-10 quoted
therein.

\bibitem{Rosato} V. Rosato, N. Pucello, G. Giuliano, {\it Evidence for cysteine
clustering in thermophilic proteomes}. { Trends in Genetics} {\bf 18}:278-281 
(2002).

\bibitem{Siggia} H.J.Bussemaker, H.Li, E.D.Siggia, {\it Building a dictionary of
genomes: identification of presumptive regulatory sites by
statistical analysis}. { Proc. Natl. Acad. Sci. USA} {\bf 97}:10096-10100 
(2000).

\bibitem{RigFlo} I. Rigoutsos, A. Floratos, 
{\it Combinatorial pattern discovery in
biological sequences: the TIREISIAS algorithm}. {Bioinformatics}
{\bf 14}:55-67 (1998).

\bibitem{Searls} D. B. Searls, {\it The Language of genes}. {Nature} {\bf
420}:211-217 (2002).

\bibitem{genebank} http://www.ncbi.nlm.nih.gov/genomes.


\bibitem{Sole} R. Ferrer Cancho, R. V. Sol\'e, {The small world of human 
language}.
{Proceedings of the Royal Society} {\bf B 268}: 2261-2266 (2001).

\bibitem{Rosato1} L. Ferraro, A. Giansanti, G. Giuliano, 
V. Rosato, in preparation.

\bibitem{note} Our definition of distance in eq.(5) looks like the 
{\it cosine coefficient}, 
an association index widely used in statistical linguistics.
See: G. Salton,{\it Automatic text processing: the trasformation, analysis and
retrieval of information by computer}. Addison-Wesley, Reading MA, (1989).

\bibitem{Huo} B. Hao, X. M.Xie, S. Y.Zang,
{\it Compositional representation of protein sequences and the
 number of Eulerian loops}. arXiv:physics/0103028 at http://lanl.arXiv.org).

\bibitem{Saitou_Nei} N. Saitou and M. Nei,
{\it The Neighbor-joining Method: A New Method for
 Recontructing Phylogenetic Trees}. {Molecular Biology and Evolution}
  {\bf 4}:406-425 (1987).

\bibitem{phylip} See the PHYLIP (the PHYLogeny Inference Package) homepage:
 http://evolution.genetics.washington.edu/phylip.html
 and the book by J. Felsenstein, {\it Inferring Phylogenies},
 Sinauer Associates, Inc. Sunderland, MA, USA (2003).

\bibitem{rdp} The Ribosomal Database Project: http://rdp.cme.msu.edu/html/

\bibitem{brown} J. R. Brown, C. J. Douady,M. J. Italia, W. E. Marshall and 
J. Stanhope, {\it Universal trees based on large combined 
protein sequence data sets}. {Nature Genetics} {\bf 28}:
281-285 (2001).

\bibitem{fitz-gibbon} C. H. House and S. T. Fitz-Gibbon, 
{\it Using Homolog Groups to Create 
a Whole-Genomic Tree of Free-living Organisms: An update}. {J. Mol. Evol.}
{\bf 54} 539-547 (2002).

\bibitem{lake1} M. C. Rivera and J. A. Lake, 
{\it The ring of life provides evidence for a 
genome fusion origin of eukaryotes}. {Nature} {\bf 431}:152-155(2004).

\bibitem{lake2} M. C. Rivera, R. Jain, J. E. Moore and J. A. Lake, 
{\it Genomic evidence for two functionally distinct gene classes}. 
{Proc. Nat. Acad. Sci. USA} {\bf 95}:6239-6244 (1998).
 
\bibitem{driskell} A. C. Driskell, C. An\'e, J. G. Burleigh, M. McMahon, 
B. C. O'Meara, M. J. Sanderson, {\it Prospects for Building the tree of 
life from large databases}. {Science} {\bf 306}:1172-1174 (2004).

\bibitem{bailin1} J. Qi, Bin Wang and Bai-Iin Hao, {\it Whole proteome 
Prokaryote Phylogeny Without Sequence Alignment: a K-string 
Composition Approach} {J. Mol. Evol.}{\bf 58 }:1-11 (2004).

\bibitem{bailin2} J. Qi, H. Luo and B. Hao, {\it CVTree: a philogenetic
tree reconstruction tool based on whole genomes } {Nucleic Acids Research}
 {\bf 32 }:W45-47 (2004). (See: http://cvtree.cbi.pku.edu.cn).

\bibitem{billera} L. J. Billera, S. P . Holmes and K. Vogtmann, 
{\it Geometry of the space of phylogenetic trees}. {Advances in Applied Math.} 
{\bf 27}:733-767 (2001). 

\bibitem{diaconis} P. Diaconis and S. Holmes, {\it Matchings and 
phylogenetic trees}. {Proc. Natl. Acad. Sci. USA} 
{\bf 95}:14600-14602 (1998).

\end{thebibliography}
\end{document}